\documentclass[twocolumn]{aastex63}
\usepackage{graphicx}
\usepackage{amsmath}
\usepackage{amssymb}
\usepackage{float}
\usepackage{subfigure}
\usepackage{lipsum}
\usepackage{gensymb}
\usepackage{mhchem}
\usepackage{graphicx}
\usepackage{mathtools}
\usepackage{scalerel}
\usepackage{times}
\usepackage{verbatim}
\usepackage{tablefootnote}
\usepackage{savesym}
\savesymbol{tablenum}
\usepackage{longtable}
\usepackage{siunitx}
\restoresymbol{SIX}{tablenum}
\usepackage{multirow}
\usepackage{url}
\usepackage{hhline}
\newif\ifAMStwofonts

\def\swift{{\it Swift}}
\def\xmm{{\it XMM-Newton}}

\def\ngc{{NGC~6814}}
\def\mrk335{{Mrk~335}}
\def\iras13{{IRAS~13224--3809}}
\def\1h07{{1H~0707--495}}
\def\izw1{{I~Zw~1}}
\def\rg{{\thinspace r_{\rm g}}}

\def\fvar{{F_{\rm var}}}

\def\feka{{Fe~K$\alpha$}}

\def\nh{{N_{\rm H}}}

\def\A{{\rm\thinspace \AA}}
\def\cm{{\rm\thinspace cm}}
\def\erg{{\rm\thinspace erg}}
\def\eV{{\rm\thinspace eV}}
\def\g{{\rm\thinspace g}}

\def\K{{\rm\thinspace K}}
\def\keV{{\rm\thinspace keV}}
\def\km{{\rm\thinspace km}}

\def\m{{\rm\thinspace m}}

\def\Msun{\hbox{$\rm\thinspace M_{\odot}$}}

\def\s{{\rm\thinspace s}}
\def\ks{{\rm\thinspace ks}}
\def\ps{{\rm\thinspace s^{-1}}}
\def\W{{\rm\thinspace W}}

\def\ergspcm{\hbox{$\erg\s\cm^{-1}\,$}}

\def\ergps{\hbox{$\erg\s^{-1}\,$}}

\def\kmps{\hbox{$\km\ps\,$}}

\def\pscm{\hbox{$\cm^{-2}\,$}}

\def\pccm{\hbox{$\cm^{-3}\,$}}

\received{xxx yyy zzz}
\revised{xxx yyy zzz}
\accepted{xxx yyy zzz}
\submitjournal{ApJL}

\shorttitle{Eclipsing the X-ray emitting region in \ngc}
\shortauthors{Gallo, Gonzalez \& Miller}

\begin{document}

\title{Eclipsing the X-ray emitting region in the active galaxy \ngc}

\correspondingauthor{Luigi C. Gallo}
\email{luigi.gallo@smu.ca}

\author{Luigi C. Gallo}
\affiliation{Department of Astronomy \& Physics, Saint Mary's University, 923 Robie Street, Halifax, Nova Scotia, B3H 3C3, Canada\\}

\author{Adam G. Gonzalez}
\affiliation{Department of Astronomy \& Physics, Saint Mary's University, 923 Robie Street, Halifax, Nova Scotia, B3H 3C3, Canada\\}

\author{Jon M. Miller}
\affiliation{Department of Astronomy, The University of Michigan, 1085 South
  University Avenue, Ann Arbor, Michigan, 48103, USA\\}



\begin{abstract}
We report the detection of a rapid occultation event in the nearby Seyfert galaxy NGC 6814, simultaneously captured in a transient light curve and spectral variability. The intensity and hardness ratio curves capture distinct ingress and egress periods that are symmetric in duration. Independent of the selected continuum model, the changes can be simply described by varying the fraction of the central engine that is covered by transiting obscuring gas.  Together, the spectral and timing analyses self-consistently reveal the properties of the obscuring gas, its location to be in the broad line region (BLR), and the size of the X-ray source to be $\sim 25\rg$.  Our results demonstrate that obscuration close to massive black holes can shape their appearance, and can be harnessed to measure the active region that surrounds the event horizon.

\end{abstract}



\section{Introduction} \label{sec:intro}
\label{sect:intro}
The innermost region around an accreting supermassive black hole produces the bulk of the radiation that defines active galactic nuclei (AGN).  This central engine is unresolved with current detectors, but simple calculations predict a light travel time of minutes-to-hours across the compact region.  

The very hot gas in this region produces X-rays, which can be used to measure the extreme relativistic effects and size scales close to the black hole \citep{1,2,3,4,5,6}.  X-ray eclipses of this region are then particularly incisive, because they can deliver clear constraints on size scales in the absence of imaging \citep{10,41,9,8,11,7}.  When size scales are known, the radiation processes that power the AGN and the nature of the inner flow are determined.  

\ngc\ ($z=0.00521$) is a Seyfert 1.5 active galaxy characterized by moderate absorption in its optical and X-ray spectrum \citep{16,17,18}.  The X-rays exhibit rapid variability on short (hours, e.g. \cite{18}) and long timescales (years, e.g. \cite{38}).  The X-ray spectrum (\cite{18,39,40}) is typical of Seyfert 1.5 AGN (e.g. \cite{40}) possessing weak excesses at low ($\lesssim1\keV$) and high ($\gtrsim10\keV$) energies, and narrow emission in the \feka\ band.

Here we report the detection of a rapid occultation event, simultaneously captured in a transient light curve and spectral variability. Together, the spectral and timing analyses self-consistently reveal the size of the X-ray source and the properties and location of the obscuring gas \citep{12,13,14,15}.

\section{Observations and Data Reduction}
\label{sect:data}
\ngc\ was observed for $\sim131\ks$ with \xmm\ \citep{19} starting 08 April, 2016.  
During the observations the EPIC detectors \citep{2a,3a} were operated in large-window mode and with the medium filter in place.  The \xmm\ Observation Data Files (ODFs) were processed to produce calibrated event lists using the \xmm\ Science Analysis System ({\sc SAS v18.0.0}).  Light curves were extracted from these event lists to search for periods of high background flaring, which were evident in the final $\sim 15\ks$ of the observation. These periods were neglected during analysis rendering a good-time exposure of $\sim 112\ks$.

Spectra were extracted from a circular region with a radius of $35$ arcsec centred on the source.  
The background photons were extracted from an off-source circular region on the same CCD with a radius of $50$ arcsec.
Pile-up was negligible during the observations.
Single and double events were selected for the pn detector, and
single-quadruple events were selected for the Metal Oxide Semi-conductor (MOS) detectors.
EPIC response matrices were generated using the {\sc SAS}
tasks {\sc ARFGEN} and {\sc RMFGEN}.  
The MOS and pn data were compared for consistency and determined to
be in agreement within known uncertainties.  The RGS \citep{33}  spectra were extracted using the {\sc SAS} task {\sc RGSPROC} and 
response matrices were generated using {\sc RGSRMFGEN}.  The combined spectra are displayed for presentation purposes.

The spectra were optimally binned \citep{5a}  and the backgrounds were modelled.
Spectral fitting was performed using {\sc XSPEC v12.9.1} \citep{6a} and fit quality is tested using the $C$-statistic \citep{7a}, since optimal binning allows bins to have a small number of counts (i.e. $<20$) and a Gaussian distribution cannot be assumed.
All parameters are reported in the rest frame of the source unless specified
otherwise, but figures remain in the observed frame.
The quoted errors on the model parameters correspond to a 90\% confidence
level for one interesting parameter.
A value for the Galactic column density toward \ngc\ of
$1.53 \times 10^{21}\pscm$ \citep{8a} is adopted in all of the
spectral fits with appropriate abundances \citep{9a}. \\

The AGN was also observed with \swift\ during this time and the XRT light curve was created with the \swift-XRT data product generator (\citealt{Evans+2009})\footnote{\url{www.swift.ac.uk/user_objects/}}. 


\section{Estimating the duration of the eclipse}
\label{sec:VAR}
\begin{figure*}
\begin{center}
\includegraphics[angle=-90,width=3.2in, trim= 3.7cm 0.cm 3cm 0.cm, clip=true]{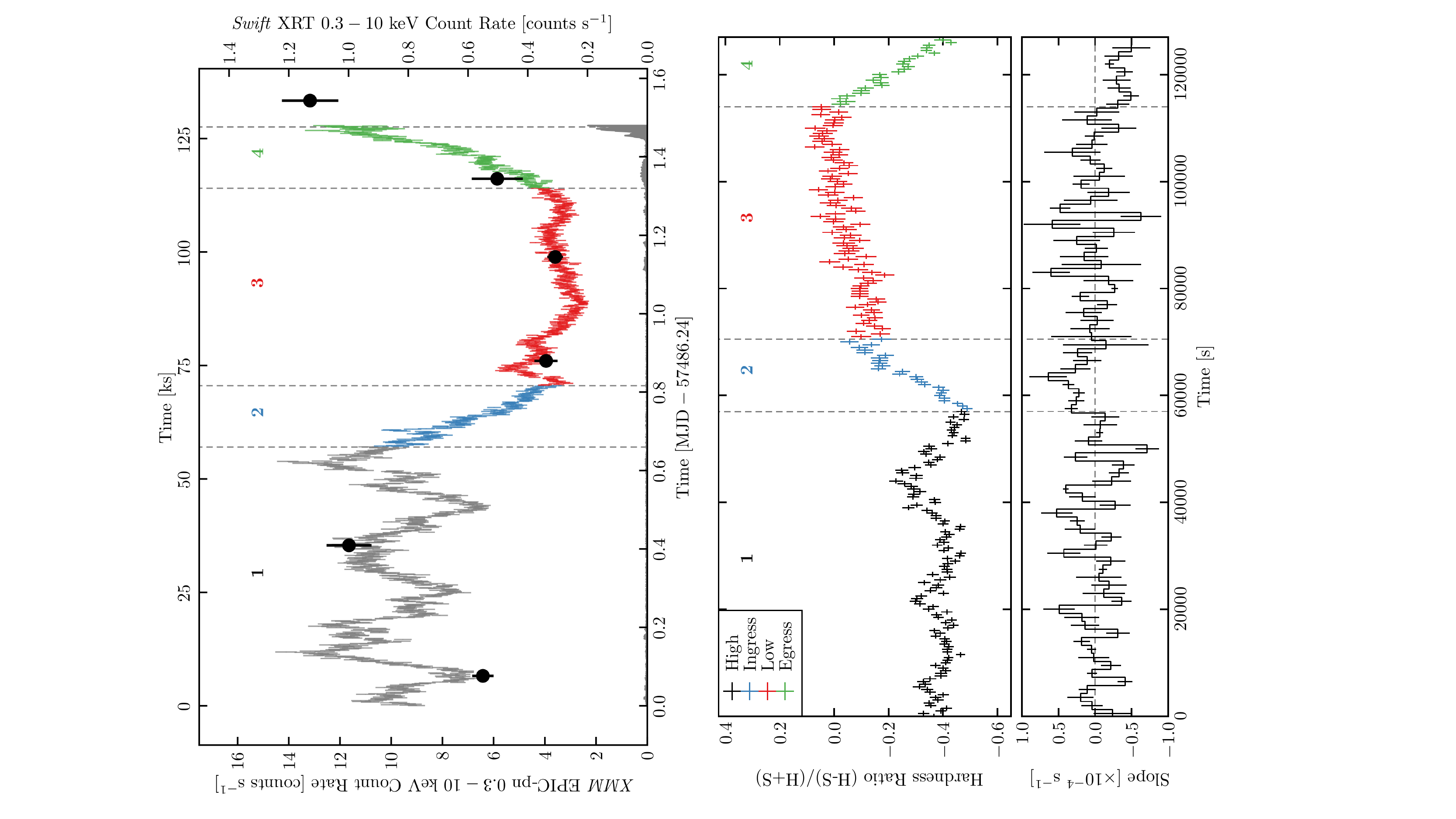}
\includegraphics[angle=-90,width=3in, trim= 1cm 2cm 1cm 1cm, clip=true]{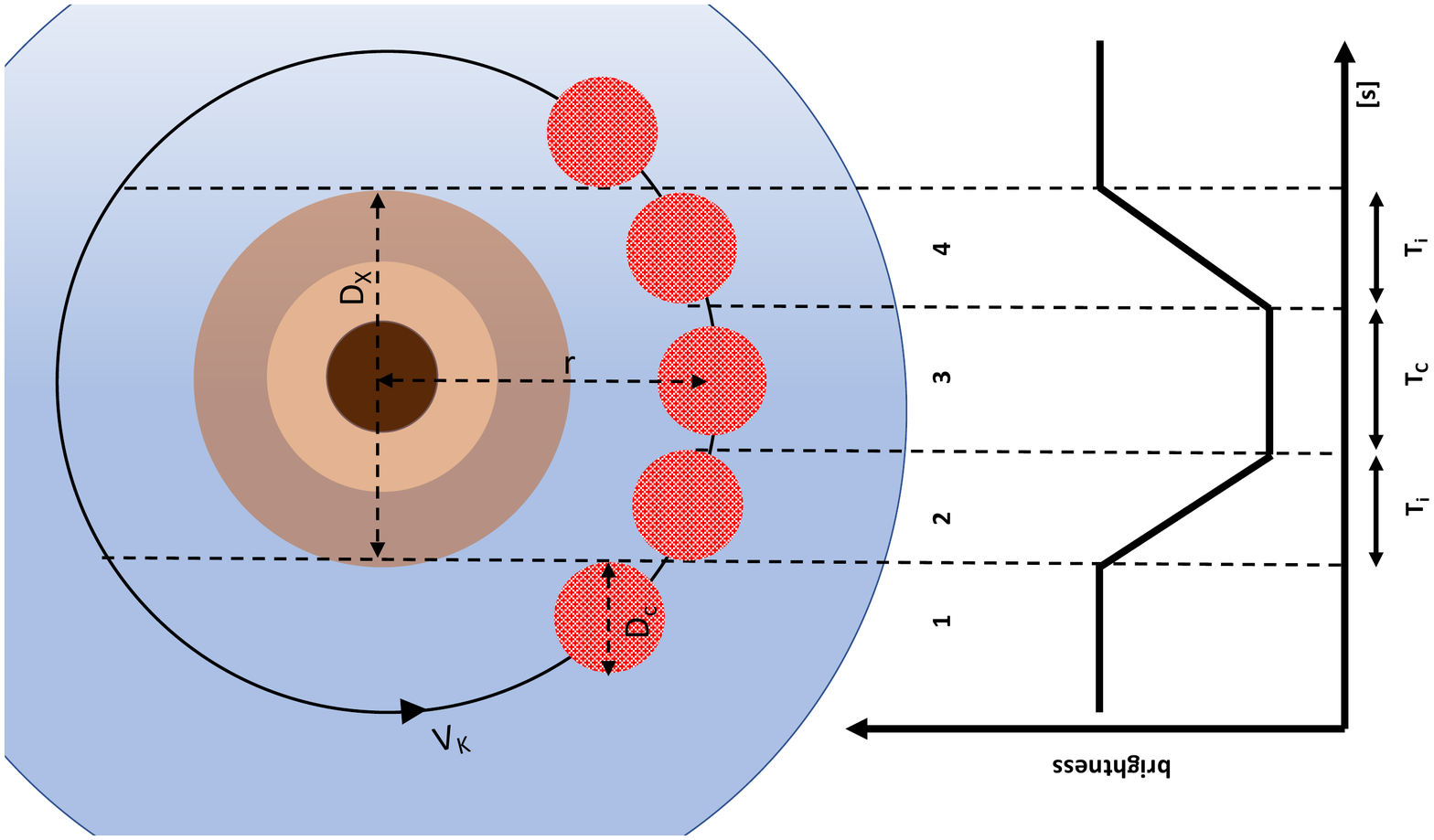}
\end{center}
\vskip -0.2in
\caption{The light curve and hardness ratio curve demonstrating the eclipsing event in \ngc. {\bf Upper left:} The \xmm\ and \swift\ (filled circles) $0.3-10\keV$ light curves of \ngc. Prior to $\sim55\ks$ demonstrates the typical random fluctuations in AGN intensity curves.  After that time, the light curve shows a constant dimming for $\sim13.5\ks$, followed by a relatively constant segment for $\sim43.5\ks$, and then brightening to its original level in $\sim13.5\ks$.  {\bf Lower left:} With the HR curve in $500~\mathrm{s}$ bins (top panel) we calculate the slope of 3 neighbouring data points before moving over one point and repeating the process until completed for the entire curve, binning the results in $1.5~\mathrm{ks}$ bins (bottom panel). Segment 1 and 3 show random changes in the slope. Segment 2 and 4 mark regions that are defined by progressive hardening and softening, respectively. We take these segments to represent ingress and egress.  {\bf Right:} Together the curves are consistent with a transient eclipsing event for which the speculative geometry is shown (the observer is viewing from the bottom of the page).
    \label{fig:lcurve}}
\end{figure*}

The $0.3-10\keV$ light curve is shown in Fig.~\ref{fig:lcurve}.  In the first $\sim60\ks$ the object exhibits the flickering behaviour that is common in AGN \citep{20}.  The light curve then depicts a steady drop in intensity followed by a relatively constant low-flux segment, ending with a steady rise to the pre-dip brightness.  This latter part of the light curve is commensurate of the transient brightness curves seen in exoplanet systems made famous with {\em Kepler} \citep{21,22}.  

Determining the start and end time of ingress and egress can be rather arbitrary based on the light curve alone. The hardness ratio (HR) curve is used to estimate these time bins more robustly. With the HR curve in $500~\mathrm{s}$ bins we calculate the slope over 3 neighbouring data points before moving over one point and repeating the process until completed for the entire curve. The results are then binned in $1.5~\mathrm{ks}$ bins (Fig.~\ref{fig:lcurve}). In this manner, we can determine at what time the slope fluctuations are no longer random. As evident in Fig.~\ref{fig:lcurve}, the HR fluctuations are random in Segment 1 and 3, but consistently hardening in Segment 2 (ingress) and softening in Segment 4 (egress). This method is completely analytical rather than relying on a by-eye approximation. 

The ingress and egress in the \ngc\ light curve are remarkably similar in duration and depth, and the transitions last for approximately $T_{\rm i}\sim13.5\ks$. During the minimum, which lasts approximately $T_{\rm C}\sim43.5\ks$, the source remains relatively hard and the brightness variations are consistent with the broadband flickering that was evident prior to ingress. \\

\section{Spectral analysis of the \xmm\ data}
\label{sec:MO}

\subsection{Spectral analysis of the RGS data}
\label{sect:NuSwiftmodel}
The RGS spectra are examined to determine the influence of the warm absorbers known to exist in this system \citep{10a,16}. Spectra are created during the high- and low-flux periods, and the two flux-resolved spectra were fitted together between $0.45-1.85~\mathrm{keV}$. Fitting the spectra with a power law plus black body to determine the continuum shape and allowing for just a change in the constant of normalization described the general shape of the spectra relatively well, but left several narrow residuals. The fit statistic was $C = 1357~\rm{for}~776~dof$. A warm absorber was then applied via the \texttt{XSPEC} implementation \citep{12a} of the \texttt{SPEX} \citep{13a} model \texttt{XABS} \citep{14a}. 

\begin{figure}
\begin{center}
\includegraphics[width=3.5in, trim= 2.cm 6.cm 1cm 6.cm, clip=true]{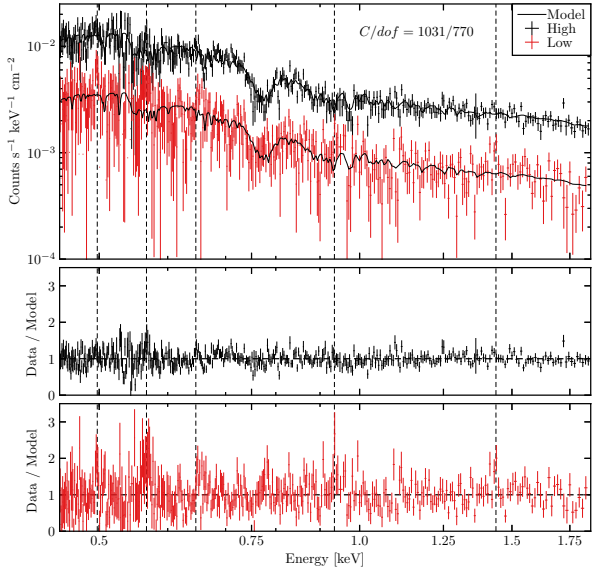}
\end{center}
\vskip -0.2in
\caption{The high-resolution RGS spectra show the presence of a warm absorber and possible emission lines during the low-flux interval.  The top panel shows the RGS spectra in the pre-eclipse high-flux state (Segment 1) and the deep minimum flux state (Segment 3).  Fitted to the data is a model including two warm absorbers (Table~\ref{tab:rgs}) and the residuals (data/model) are shown in the next two panels.  The low-flux level can be well described by simply renomalizing the high-flux model.  However possible emission-like residuals are evident in the low-flux state (lower panel) at $\sim0.50~\mathrm{keV}$, $0.57~\mathrm{keV}$, $0.65~\mathrm{keV}$, $0.94~\mathrm{keV}$, and $1.445~\mathrm{keV}$, with each line improving the fit by $\Delta C > 10$ for $3$ additional free parameters
    \label{fig:rgs}}
\end{figure}

The addition of one warm absorber with free column density ($\nh$), ionization parameter ($\xi= L/nr^2$, where n is the density of the cloud at a distance r from the source of ionising luminosity L), and redshift changed the statistic by $\Delta C = 240$ for 3 additional parameters. The residuals were improved, but significant absorption-like residuals remained around $0.75~\mathrm{keV}$ and $0.95~\mathrm{keV}$. The addition of a second warm absorber produced a good quality fit without any substantial deviations in the residuals. The final fit to the flux resolved spectra is $C = 1031~\rm{for}~770~dof$. The model parameters are shown in Table~\ref{tab:rgs} and the fit residuals are shown in Fig.~\ref{fig:rgs}. These warm absorbers were applied to the EPIC model for the broadband spectrum.

The low-flux spectrum displays several excess deviations from the described model, exhibiting emission lines not present in the high-flux data. A search for significant positive residuals (i.e. stepping a Gaussian through $0.45-1.85~\mathrm{keV}$ every $5~\mathrm{eV}$) results in 5 possible features (Fig.~\ref{fig:rgs}) at $0.50~\mathrm{keV}$, $0.57~\mathrm{keV}$, $0.65~\mathrm{keV}$, $0.94~\mathrm{keV}$, and $1.445~\mathrm{keV}$, with each line improving the fit by $\Delta C > 10$ for $3$ additional free parameters. \\

\subsection{The EPIC-pn spectra}
\label{sect:NuSwiftmodel}
\begin{figure}
\begin{center}
\includegraphics[angle=0,width=3.5in, trim= 2cm 8cm 2cm 8cm, clip=true]{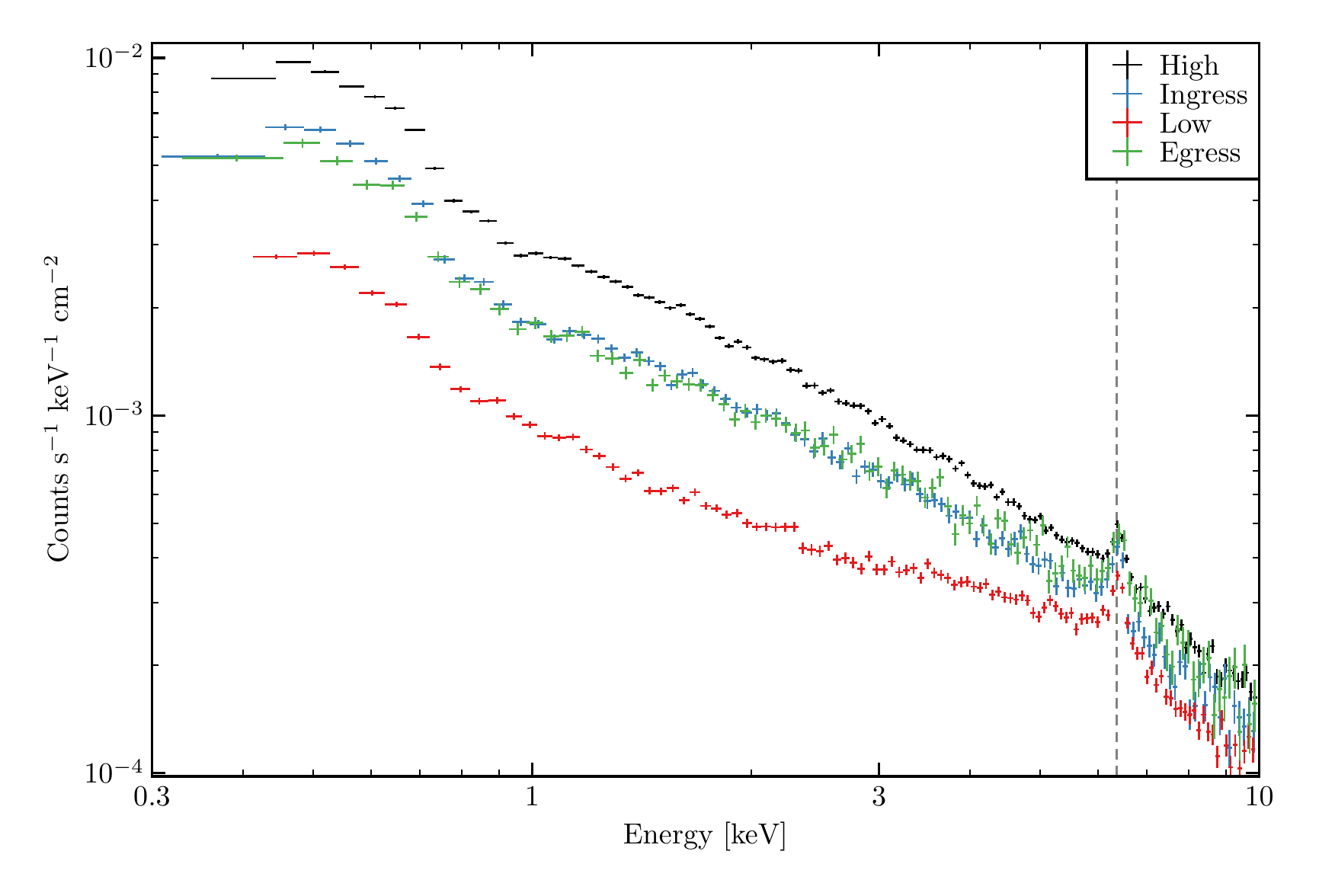}
\includegraphics[width=3.5in, trim= 2cm 8cm 2cm 8.5cm, clip=true]{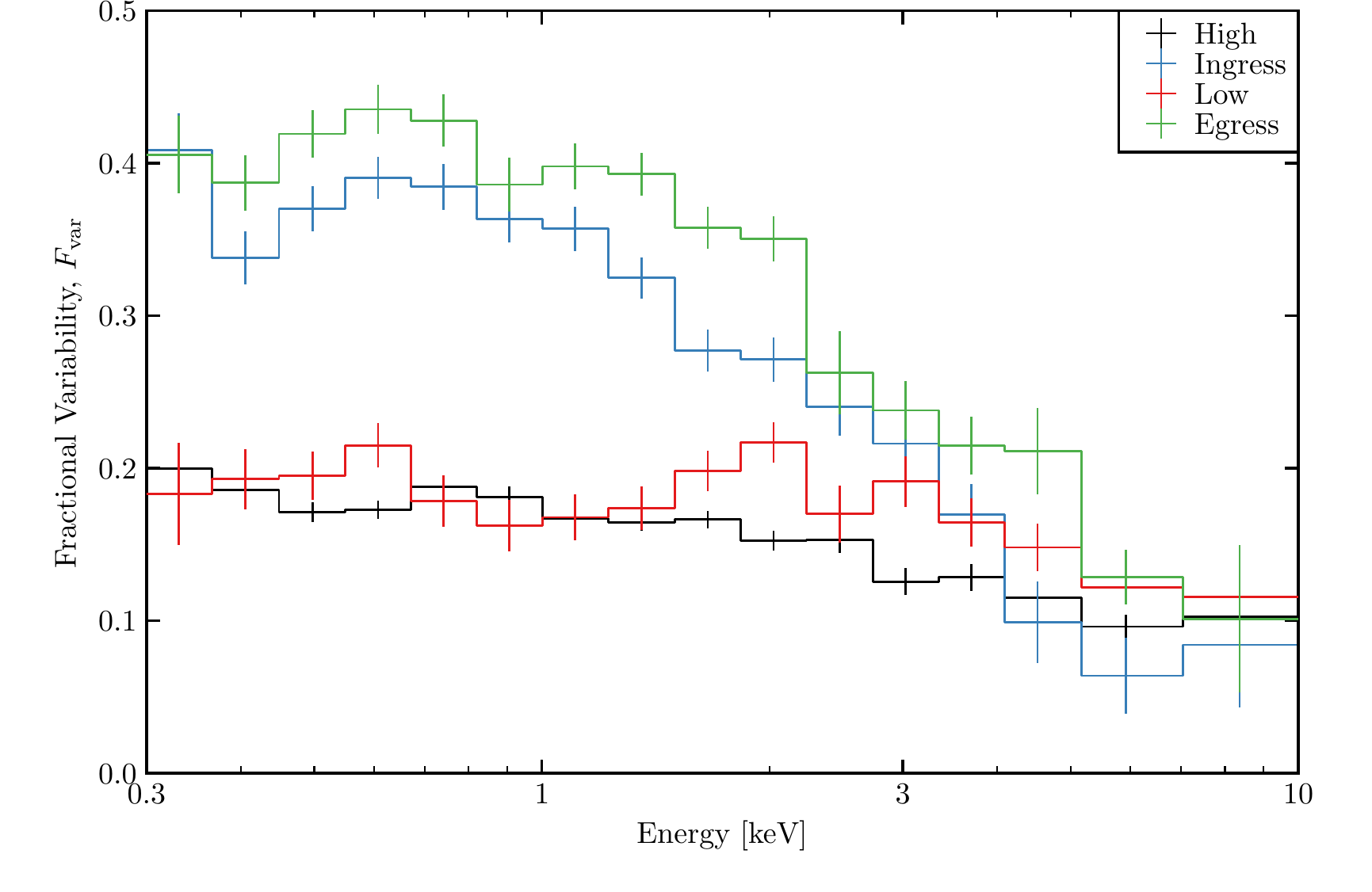}
\includegraphics[width=3.3in, trim= 1cm 1.5cm 2cm 2cm, clip=true]{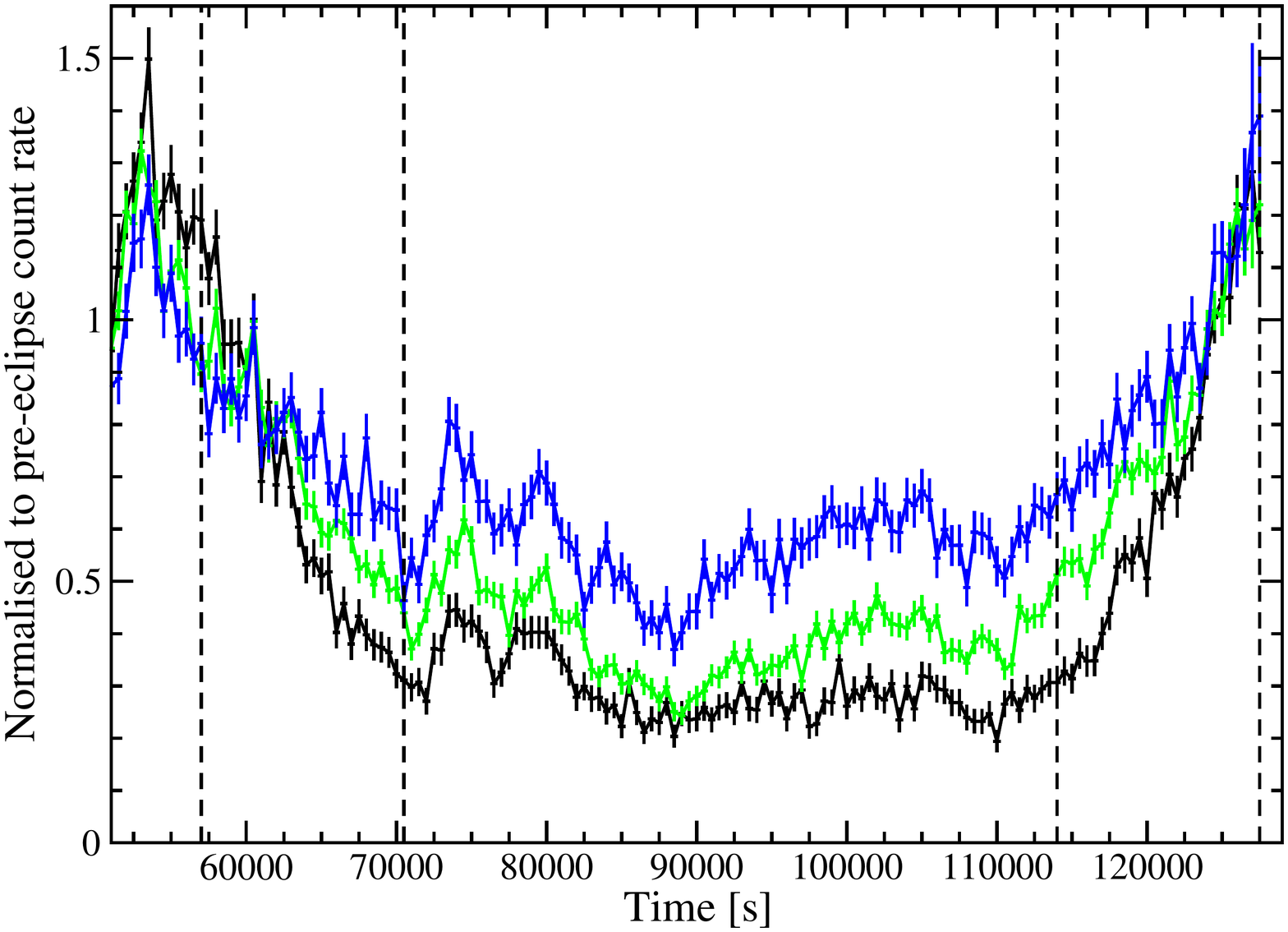}
\end{center}
\vskip -0.2in
\caption{The spectral changes show the effects of the transient absorber and energy dependence on the eclipse. {\bf Top:} The EPIC-pn spectra are created in the four time segments shown in Fig.~\ref{fig:lcurve}.  The spectra are remarkably similar during ingress and egress, and clearly harder when the source is dimmer.  {\bf Middle:} The fractional variability spectra show the degree of variability during the different segments. During ingress and egress, the variability is clearly dominated by the low-energy emission, which would be consistent with increasing absorption.  During the high- and low-flux states, the variability is similar suggesting that the nature of the fluctuations are probably alike in the high and low-flux intervals (i.e. the intrinsic nature has not changed).  Above $\sim5\keV$, the variability is similar at all flux levels indicating the eclipse has little effect at these highest energies. {\bf Bottom:} Light curves in the $0.6-0.8\keV$ (black), $2-4\keV$ (green), $4-6\keV$ (blue) show the depth of the eclipse is energy dependent as would be expected because of the column density of the cloud.  There is indication the shape of the transient may also differ as a function of energy (e.g. see the transitions in the ingress segment).
    \label{fig:fvar}}
\end{figure}

\begin{figure*}
\begin{center}
\includegraphics[width=6in,trim= 0cm 5cm 0cm 5cm, clip=true]{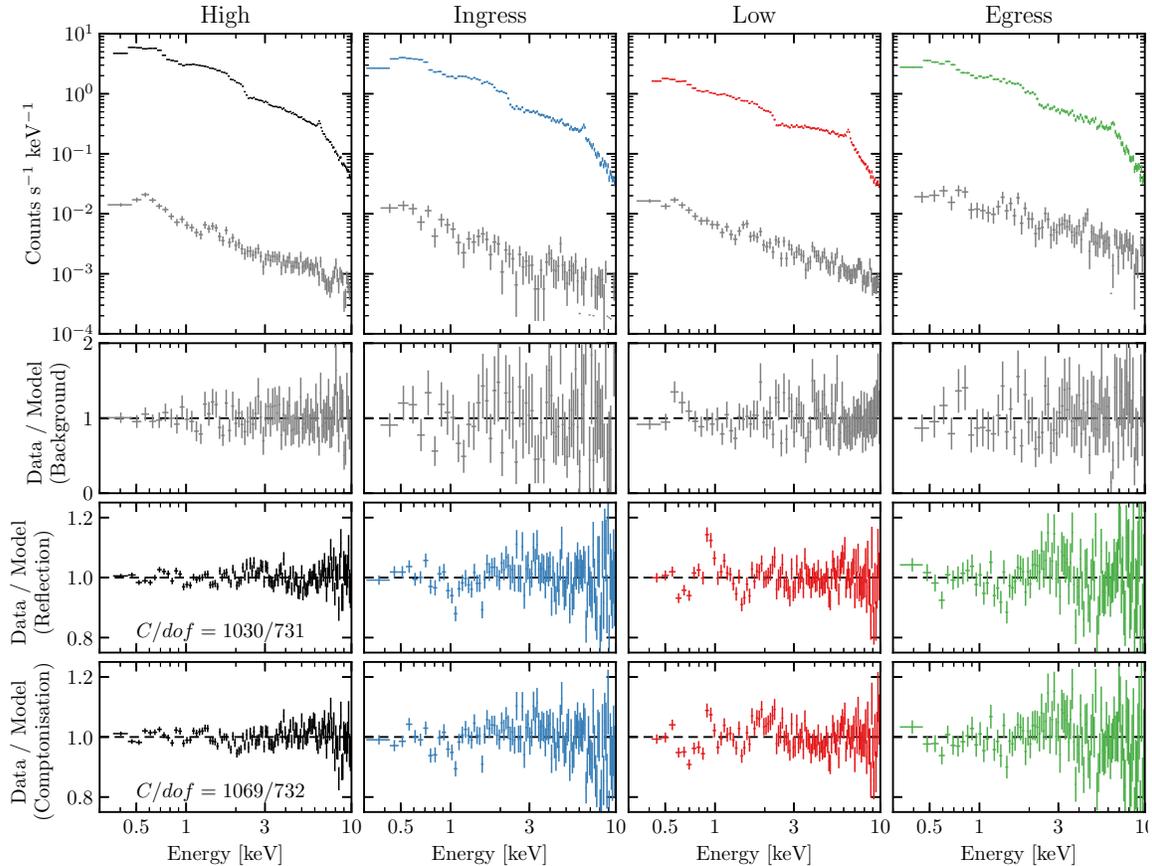}
\end{center}
\vskip -0.2in
\caption{Partial covering models applied to different continuum scenarios.  {\bf Top row: }The source and background pn spectra between $0.3-10\keV$ during each segment (labeled on top).  {\bf Upper middle row:} The residuals from fitting the background data.  {\bf Lower middle row:} The resulting residuals from the ionised blurred reflection model described in the text and in Table~\ref{tab:comp}.  {\bf Bottom row:} The resulting residuals from the soft Comptonization model described in the text and in Table~\ref{tab:comp}.
    \label{fig:pn}}
\end{figure*}

EPIC-pn spectra were created in each of the four time segments representing the pre-eclipse, ingress, low-state, and egress (Fig.~\ref{fig:fvar} and ~\ref{fig:pn}).  The spectra corroborate the behaviour in the hardness ratio and light curves.  The fractional variability spectrum ($\fvar$, Fig.~\ref{fig:fvar}), which illustrates the level of variability in each energy band \citep{23}, confirms the variations during ingress and egress are predominately at lower energies, and more achromatic during the minimum and pre-ingress periods.  Moreover, the variability above $\sim6\keV$ is comparable during the entire observation (Fig.~\ref{fig:fvar}).

For the simplest model, we fitted the spectra with a single power law plus fixed warm absorbers, modified by a partial coverer.  Even with the power law and partial coverer components permitted to vary, this resulted in a relatively poor fit ($C = 2008~\rm{for}~728~dof$), and demonstrated the need for more physical continuum models.  Notwithstanding this, it was notable that the effects of the partial coverer was more enhanced in the low-state than during pre-eclipse.

The continuum was modelled assuming the blurred reflection scenario \citep{24}.  Here, some fraction of the corona illuminates the inner accretion disc producing backscattered emission that is modified for Doppler, Special, and General relativistic effects.  We use {\sc relxillD} \citep{16a,17a} to model the scenario, which allows the density of the disc to be altered.  

The spectra are modified by the double warm absorber system that was found in the RGS analysis (Fig.~\ref{fig:rgs}).  We determined that refitting the warm absorbers to the pn data did not provide a substantially better fit than simply using the RGS measured parameters.  Therefore, the warm absorber parameters were fixed to the RGS values throughout the analysis.  An \feka\ emission line was present in the data, but at energies slightly higher than $6.4\keV$.  A simple Gaussian profile is used to model this component.  The line energy and width were linked between the epochs, but the normalisation was left free to examine for variability.  

Attempting to fit all four spectra in a self-consistent manner, we allow only the photon index and power law flux to vary between epochs, which are expected to vary on such time scales.  The reflection fraction ($\mathcal R$) was linked indicating that the ratio of reflected-to-continuum flux was not changing.  This could be examined in future modelling, but the rather constant $\fvar$ spectra (Fig.~\ref{fig:fvar})  suggests modest spectral variability during the high- and low-flux intervals.  This model resulted in a rather poor fit to the data with $C = 2198~\rm{for}~737~dof$.

A partial coverer was applied to the central X-ray emission component only (i.e. {\sc relxillD}).  This was a marked improvement to the fit.  The best fit was obtained when  the partial coverer had a fixed column density and ionisation parameter, but the covering fraction ($C_f$) was permitted to vary between epochs (Fig.~\ref{fig:pn} and Table~\ref{tab:comp}).  In this manner, the covering fraction changed from $C_f \approx 0$ prior to the eclipse and $C_f \approx 0.56$ in the minimum flux state.  The final fit statistic was $C = 1029~\rm{for}~731~dof$

Allowing the partial coverer to be neutral rather than ionised resulted in a poorer fit ($\Delta C = 22$ for 1 fewer parameter).  Allowing the column density to vary between epochs rather than the covering fraction also resulted in a poorer fit ($C = 1537~\rm{for}~731~dof$).

We considered if the partial covering results could be dependent on the assumed X-ray continuum.  To test this we replaced the blurred reflection model with a soft-Comptonisation model for the continuum.  This is another commonly used X-ray model that attributes the soft-excess emission to an optically thick warm corona that is situated over the disc \citep{18a,19a,20a}.  We adopt the model previous used in \cite{21a}, which incorporates two {\sc nthcomp} components, one for the traditional optically-thin, hot corona and the other for the warm corona.  As with the blurred reflection model, the continuum is modified by warm absorbers and a $\sim 6.45\keV$ Gaussian profile is included.  Various combinations could be attempted to describe the intrinsic variability.  We found the the simplest scenario was to link the warm corona parameters, but allow the hot corona to vary from segment-to-segment.  

A partial coverer is again added and the covering fraction is free to vary (Fig.~\ref{fig:pn} and Table~\ref{tab:comp}) .  The partial coverer column density and ionisation parameter are very similar to what was found with the blurred reflection continuum.  The covering fractions were offset by about $+20$ per cent at each interval compared to the reflection model, but the relative change between intervals was the same as in the reflection model.  Despite using a different continuum model, the partial covering parameters were very similar.  The best-fit soft-Comptonisation model resulted in a fit statistic of $C = 1131~\rm{for}~728~dof$

\begin{table}
	\centering
	\caption {The RGS data fitted with a double warm absorber and a phenomenological continuum of a blackbody and power law. In {\sc xspec } terminology: \texttt{tbabs} $\times$ \texttt{WA1} $\times$ \texttt{WA2} $\times~($ \texttt{bb}$~+~$\texttt{po} $)$.}
	\label{tab:rgs}
	\bigskip
	\sffamily
	\begin{tabular}{ccc}
		\hline
		\hline
		Model Component & Model Parameter & Parameter Value \\
		\hline
		\texttt{xabs}$_{1}$ & $\log\xi / \mathrm{erg~cm^{-2}~s^{-1}}$ & $2.89^{+0.11}_{-0.12}$ \\
		& $N_{\mathrm{H}} / 10^{21}~\mathrm{cm^{-2}}$ & $36.8^{+16.7}_{-15.5}$ \\
		& $v_{\mathrm{out}} / \mathrm{km~s^{-1}}$ & $5333^{+262}_{-380}$ \\
		\hline
		\texttt{xabs}$_{2}$ & $\log\xi / \mathrm{erg~cm^{-2}~s^{-1}}$ & $0.96^{+0.10}_{-0.07}$ \\
		& $N_{\mathrm{H}} / 10^{21}~\mathrm{cm^{-2}}$ & $3.85\pm0.72$ \\
		& $v_{\mathrm{out}} / \mathrm{km~s^{-1}}$ & $5569^{+245}_{-224}$ \\
		\hline
		\texttt{bb} & $kT_{\mathrm{e}} / \mathrm{eV}$ & $111\pm5$ \\
		& $\mathrm{Norm} / 10^{-4}$ & $2.4\pm0.3$ \\
		\hline
		\texttt{po} & $\Gamma$ & $1.5\pm0.2$ \\
		& $\mathrm{Norm} / 10^{-3}$ & $7\pm1$ \\
		\hline
		\texttt{const} & $F_{\mathrm{low}} / F_{\mathrm{high}}$ & $0.276\pm0.009$ \\
		\hline
		\hline
	\end{tabular}
\end{table}

\begin{table*}
	\centering
	\caption{The partial covering model applied to two different continuum scenarios, Comptonization and blurred reflection.  In {\sc xspec } terminology the Comptonization model is: \texttt{tbabs} $\times$ \texttt{ztbabs} $\times$ \texttt{WA1} $\times$ \texttt{WA2} $\times~($ \texttt{zxipcf} $\times~($ \texttt{nthcomp}$_{\mathrm{soft}}~+~$\texttt{nthcomp}$_{\mathrm{hard}})~+~$\texttt{zgauss} $)$ and the blurred reflection model is: \texttt{tbabs} $\times$ \texttt{WA1} $\times$ \texttt{WA2} $\times~($ \texttt{zxipcf} $\times~($ \texttt{cflux} $\times$ \texttt{cutoffpl} $+$ \texttt{const} $\times$ \texttt{cflux} $\times$ \texttt{relxillD} $)~+~$\texttt{zgauss} $)$. Parameter values without uncertainties are fixed during the fitting.}
	\label{tab:comp}
	\bigskip
	\scriptsize
	\sffamily
	\begin{tabular}{cccccccc}
		\hline
		\hline
	Continuum	& Model  & Model  & \multicolumn{5}{c}{Parameter Value} \\
	 Model & Component	& Parameter & All & High & Ingress & Low & Egress \\
		\hline
Comptonization &		\texttt{ztbabs} & $N_{\mathrm{H}} / 10^{20}~\mathrm{cm^{-2}}$ & $5.95^{+2.37}_{-1.69}$ & & & & \\
		\hline
		& \texttt{zxipcf} & $N_{\mathrm{H}} / 10^{23}~\mathrm{cm^{-2}}$ & $1.37^{+0.07}_{-0.04}$ & & & & \\
		&				& $\log\xi / \mathrm{erg~cm^{-2}~s^{-1}}$ & $1.09^{+0.05}_{-0.03}$ & & & & \\
		&				& $f_{\mathrm{cov}}$ & & $0.16^{+0.04}_{-0.06}$ & $0.44^{+0.02}_{-0.04}$ & $0.80\pm0.01$ & $0.52^{+0.03}_{-0.04}$ \\
		\hline
	&	\texttt{nthcomp}$_{\mathrm{soft}}$ & $\Gamma$ & $2.10^{+0.36}_{-0.24}$ & & & & \\
	&									   & $kT_{\mathrm{e}} / \mathrm{eV}$ & $113^{+10}_{-3}$ & & & & \\
	&									   & $kT_{\mathrm{bb}} / \mathrm{eV}$ & $3$ & & & & \\
	&									   & $\mathrm{Norm} / 10^{-3}~\mathrm{cm^{-2}~s^{-1}~{keV}^{-1}}$ & $1.63^{+0.24}_{-0.17}$ & & & & \\
		\hline
	&	\texttt{nthcomp}$_{\mathrm{hard}}$ & $\Gamma$ & & $1.70\pm0.03$ & $1.74^{+0.02}_{-0.04}$ & $1.97\pm0.04$ & $1.77\pm0.04$ \\
	&									   & $kT_{\mathrm{e}} / \mathrm{keV}$ & $100$ & & & & \\
	&									   & $kT_{\mathrm{bb}} / \mathrm{eV}$ & $3$ & & & & \\
	&									   & $\mathrm{Norm} / 10^{-3}~\mathrm{~\mathrm{cm^{-2}~s^{-1}~{keV}^{-1}}}$ & & $8.61^{+0.52}_{-0.68}$ & $8.58^{+0.70}_{-0.76}$ & $12.10\pm1.00$ & $9.48\pm0.90$ \\
		\hline
	&	\texttt{zgauss} & $E / \mathrm{keV}$ & $6.45\pm0.01$ & & & & \\
	&					& $\sigma / \mathrm{eV}$ & $141^{+24}_{-19}$ & & & & \\
	&					& $\mathrm{Norm} / 10^{-5}~\mathrm{ph.~cm^{-2}~s^{-1}}$ & & $6.03^{+0.78}_{-0.60}$ & $5.24^{+1.01}_{-1.05}$ & $5.09^{+0.72}_{-0.58}$ & $6.45^{+1.54}_{-1.66}$ \\
		\hline
		\hline
		Blurred reflection & \texttt{zxipcf} & $N_{\mathrm{H}} / 10^{23}~\mathrm{cm^{-2}}$ & $1.11\pm0.12$ & & & & \\
					&	& $\log\xi / \mathrm{erg~cm^{-2}~s^{-1}}$ & $1.09^{+0.03}_{-0.16}$ & & & & \\
					&	& $f_{\mathrm{cov}}$ & & $<0.01$ & $0.20\pm0.03$ & $0.56\pm0.02$ & $0.29\pm0.04$ \\
		\hline
		& \texttt{cflux} & $\log F / \mathrm{erg~cm^{-2}~s^{-1}}$ & & $-10.105^{+0.006}_{-0.013}$ & $-10.187^{+0.007}_{-0.020}$ & $-10.247^{+0.010}_{-0.020}$ & $-10.135^{+0.011}_{-0.024}$ \\
		& \texttt{cutoffpl} & $\Gamma$ & & $1.99^{+0.02}_{-0.01}$ & $1.98^{+0.02}_{-0.03}$ & $1.95^{+0.01}_{-0.02}$ & $1.93^{+0.02}_{-0.03}$ \\
		\hline
		& \texttt{const} & $\mathcal R$ & $1.16^{+0.12}_{-0.06}$ & & & & \\
		& \texttt{relxillD} & $q_{\mathrm{in}}$ & $8.48^{+0.84}_{-0.56}$ & & & & \\
		& 			 & $q_{\mathrm{out}}$ & $3$ & & & & \\
		& 			 & $R_{\mathrm{b}} / \rg$ & $6$ & & & & \\
		& 			 & $a/{\mathrm{[cJ/GM^2]}}$ & $0.998$ & & & & \\
 		&				  & $i / ^{\circ}$ & $66^{+1}_{-4}$ & & & & \\
		&				  & $\log\xi / \mathrm{erg~cm^{-2}~s^{-1}}$ & $0.38^{+0.05}_{-0.07}$ & & & & \\
		&				  & $\log N_{\mathrm{H}} /~\mathrm{cm^{-2}}$ & 19 & & & \\
		&				  & $A_{\mathrm{Fe}}$ & $3.9^{+0.7}_{-0.9}$ & & & & \\
		\hline
		& \texttt{zgauss} & $E / \mathrm{keV}$ & $6.45\pm0.01$ & & & & \\
		&				& $\sigma / \mathrm{eV}$ & $137^{+21}_{-16}$ & & & & \\
		&				& $\mathrm{Norm} / 10^{-5}~\mathrm{ph.~cm^{-2}~s^{-1}}$ & & $5.41^{+0.79}_{-0.56}$ & $5.47^{+0.96}_{-0.86}$ & $5.27^{+0.64}_{-0.48}$ & $6.77^{+1.39}_{-1.26}$ \\
		\hline
		\hline

	\end{tabular}
\end{table*}

\section{Discussion and Conclusions}
\label{sec:Disc}

Similar to exoplanet light curves, the transient light curve in \ngc\ can be interpreted as an occultation event, in this case, of the primary X-ray source by an orbiting globule.  Such events have been reported previously \citep{10,9,11}, but this may be the first time a rapid occultation is captured in its entirety with spectral and temporal data.  The symmetry in the transient light curve indicates the obscurer is rather uniform and possibly a single cloud.  The rapid time scales of the eclipse place the cloud close to the black hole, and the fact the dip does not reach zero brightness implies the obscurer only partially covers the X-ray source.   

The illustration in Fig.~\ref{fig:lcurve} highlights the situation and demonstrates the parameters that can be estimated from the spectral and temporal measurements.  The depth of the eclipse is energy-dependent (Fig.~\ref{fig:fvar}), with the low-energy X-rays diminishing to $\sim20$ per cent of the pre-eclipse brightness and the high-energy X-rays to $\sim50$ per cent.  There is some indication the shape of the eclipse may also differ -- at higher energies, the transitions between time points might be smoother than at lower energies.  This may be an indication the source size is energy dependent and that the high-energy X-rays are more centrally compact.  Such limb darkening behaviour is common in stellar and exoplanet transient curves.

We modeled the $0.3-10\keV$ spectra from each of the time segments (see Fig.~\ref{fig:lcurve}) simultaneously in a self-consistent manner.  We tested different continuum models, which yielded similar results, but for ease of presentation, here we will discuss the results assuming the intrinsic X-ray emission is described by ionised blurred reflection \citep{24}.  The continuum emission was also modified by two non-variable ionized (warm) absorbers and a Gaussian profile at $6.45\pm0.01\keV$.  This primary X-ray emission was then obscured by a partial coverer \citep{25,26} that was of constant column density and ionization parameter. The normalization (brightness) and photon index ($\Gamma$) of the power law continuum were free to vary between segments as is typical in Seyfert galaxies.  The reflection fraction ($\mathcal{R}$) was linked indicating the relative fraction of reflected-to-continuum emission remained constant.  For the partial coverer, only the covering fraction ($C_f$) varied between the segments.  This fit was acceptable yielding a C-statistic of $C = 1029~ \rm{for}~ 731$ degrees of freedom.  The partial coverer could be described as having $\nh=(11.1\pm1.2) \times 10^{22} \pscm$ and $\xi = (12.3^{+0.9}_{-3.8}) \ergspcm$.  Prior to ingress (Segment 1), the covering fraction was $C_f <0.01$.  During ingress (Segment 2) and egress (Segment 4), $C_f = 0.20\pm 0.03$ and $0.29\pm 0.04$, respectively.  The maximum covering fraction ($C_f = 0.56\pm 0.02$) was returned during the minimum flux (Segment 3).  

The mass of the black hole in \ngc\ is $M_{BH} = 1.09 \times 10^7 \Msun$ \citep{27}.  Following \citep{11} (section 6), assuming the obscurer is moving on a Keplerian orbit there is a relationship between the orbit and obscurer properties such that:
$r^{5/2} = (GM)^{1/2} \frac{L_X \Delta T}{N_{H} \xi}$.
The total duration of the eclipse (from onset of ingress to end of egress) is $\Delta T\approx70.5\ks$ and the X-ray luminosity prior to eclipse is measured to be $L_X \approx 2\times10^{42}\ergps$.  This would place the partial coverer at $r\approx 2694\rg = 4.34\times 10^{15} \cm$.

The electron density of the cloud is $n = L_X /r^{2}\xi =8.61^{+0.63}_{-2.65}\times 10^{9} \pccm$ and from the measured column density we obtain a cloud diameter of approximately $D_{\rm C}=1.30^{+0.17}_{-0.42}\times 10^{13} \cm$.   The Keplerian velocity of the partial coverer is $V_{\rm K} = D_{\rm C} / T_{\rm i} \approx 10\times10^{3}\kmps$.  The density is comparable to the electron densities estimated for broad-line region (BLR) ``clouds'' and the velocity and distance are also in agreement with the inner BLR \citep{28,29,30}.  

The narrow \feka\ emission line in \ngc\ exhibits a relatively constant flux within uncertainties ($\sim \pm15$ per cent) in the four different segments implying that it is not affected by the partial coverer.  Its equivalent width does change in accordance with continuum flux changes (i.e. largest equivalent width during the low-flux interval). We  modelled  the line with a Gaussian profile and found it at $E=6.45\pm0.01\keV$ and with $\sigma=136^{+21}_{-16}\eV$. The resulting $FWHM=320^{+49}_{-38}\eV$, which renders a velocity of about $(15\pm2)\times10^{3}\kmps$.  This is compatible with the location of the obscurer so we may have an example of obscuration and re-emission from the same region. 

Comparing the duration of ingress to the duration of the low-flux interval provides an estimate of the X-ray source size: $D_{\rm X} = D_{\rm C} \times T_{\rm C}/ T_{\rm i} = 4.20^{+0.54}_{-1.37}\times 10^{13} \cm = 26^{+3}_{-8} \rg$.  The value is completely consistent with what is normally estimated or assumed for the size of the corona ($\sim 20\rg$) \citep{2,4,6,31,32,41}.

There are other interesting aspects of the eclipse that are observed.  The RGS spectra are generated for the high- and low-flux intervals (Segment 1 and 3, respectively).  The high-resolution grating data are well fitted with two warm absorbers that are then applied to the CCD spectra.  The difference between the low- and high-flux states can largely be attributed to a change in normalization, but there are narrow emission features that appear during eclipse.  These could be attributed to some scattered emission from the partial coverer or from emission that originates at large distances from the black hole that is only evident when the continuum brightness is suppressed (e.g. in the narrow-line region,  starburst region, or torus) \citep{34,35,36}. 

The detection of rapid eclipsing events in AGN light curves are powerful tools for determining properties of the absorber, the BLR, and of the primary X-ray source. Here, we have reported the discovery of what appears to be a single cloud passing in front of the central engine. The results show that relativistic reflection and partial covering are both natural to the accretion flow and necessary for accurate modeling, not effects that naturally oppose each other. The properties of the absorber imply a BLR origin.  The size of the X-ray source is compact and consistent with expectations \citep{2,4,6,31,32,41}.  The eclipse shows evidence of energy-dependent effects, which may lead to understanding limb darkening in the corona.

Such events are difficult to detect in the stochastic behaviour of AGN light curves, but they are probably not rare.    At least some Seyfert 1.5s may offer advantageous viewing angles: high enough to intercept the BLR in a manner that can give eclipses, but low enough to avoid being blocked by a torus.  \ngc\ is probably an excellent target for witnessing such an event.  The designation of \ngc\ as an intermediate Seyfert~1.5 implies we are seeing the AGN at higher inclinations (i.e. more edge-on).  Estimates place the BLR in \ngc\ at an inclination of $60-80$ degrees \citep{17}, which is consistent with our X-ray measured disc inclination ($67^{+1}_{-4}$~degrees).  From our line-of-sight, the BLR crosses the X-ray source, but through relatively modest obscuration from the torus.  

Data in current X-ray archives can be used to search for similar episodes.  Long, uninterrupted observations of well-selected sources can be studied more extensively using current missions, and studied even better with future missions.

\section*{Acknowledgements}
We thank the referee for providing a helpful report.  The \xmm\ project is an ESA Science Mission with instruments and contributions directly funded by ESA Member States and the USA (NASA). This work made use of data supplied by the UK Swift Science Data Centre at the University of Leicester, data from the NuSTAR mission, a project led by the California Institute of Technology, managed by the Jet Propulsion Laboratory, and funded by the National Aeronautics and Space Administration. L.C.G and A.G.G. are support by NSERC and the CSA.  J.M.M. is supported by NASA
funding, through \emph{Chandra} and \emph{XMM-Newton} guest observer
programs.    


\bibliography{bibfile}
\bibliographystyle{aasjournal}
\end{document}